\documentclass[longauth]{aa}  

\usepackage{graphicx}
\usepackage{txfonts}
\usepackage[english]{babel}
\usepackage[hidelinks]{hyperref}
\usepackage[symbol]{footmisc}

\bibpunct{(}{)}{;}{a}{}{,} 
\begin{document} 

   \title{Particle monitoring capability of the Solar Orbiter Metis coronagraph through the increasing phase of solar cycle 25}
\titlerunning{Particle monitoring with Metis on Solar Orbiter}

   

   \author{C. Grimani \inst{1,2}
          \and V. Andretta \inst{3} 
          \and E. Antonucci \inst{4}
          \and P. Chioetto \inst{5,6}
          \and V. Da Deppo \inst{5} 
          \and M. Fabi\inst{1,2}
          \and S. Gissot \inst{7} 
          \and G. Jerse \inst{8}
          \and M. Messerotti \inst{8}
          \and G. Naletto \inst{9,5}
          \and M. Pancrazzi \inst{4}
          \and A. Persici\inst{10}
          \and C. Plainaki \inst{11}
          \and M. Romoli \inst{12,13}
          \and F. Sabbatini \inst{1,2}
          \and D. Spadaro \inst{14}
          \and M. Stangalini \inst{11}
          \and D. Telloni \inst{4}
          \and L. Teriaca \inst{15}
          \and M. Uslenghi \inst{16}
          \and M. Villani \inst{1,2}
            \and L. Abbo \inst{4}
          \and A. Burtovoi \inst{13}
          \and F. Frassati \inst{4}
          \and F. Landini \inst{4}
          \and G. Nicolini \inst{4}
          \and G. Russano \inst{3}
          \and C. Sasso \inst{3}
          \and R. Susino \inst{4}
          }

   \institute{DiSPeA, University of Urbino Carlo Bo, Urbino (PU), Italy\\
        \email{catia.grimani@uniurb.it}
    \and{INFN, Florence, Italy}
    \and{INAF – Astronomical Observatory of Capodimonte, Naples, Italy}
    \and{INAF – Astrophysical Observatory of Torino, Italy}
    \and{CNR – IFN, Via Trasea 7, 35131, Padova, Italy}
    \and{CISAS, Centro di Ateneo di Studi e Attivit\`a Spaziali "Giuseppe Colombo", via Venezia 15, 35131 Padova, Italy}
    \and{Solar-Terrestrial Centre of Excellence – SIDC, Royal Observatory of Belgium, Ringlaan -3- Av. Circulaire, 1180 Brussels, Belgium}
    \and{INAF - Astrophysical Observatory of Trieste, Italy}
    \and{Dip. di Fisica e Astronomia “Galileo Galilei”, Università di Padova, Via G. Marzolo, 8, 35131, Padova Italy}
    \and{Politecnico di Milano, Milano, Italy}
    \and{ASI – Italian Space Agency, Via del Politecnico snc, 00133 Rome, Italy}
    \and{University of Florence, Physics and Astronomy Department, Largo E. Fermi 2, 50125 Florence, Italy}
    \and{INAF – Arcetri Astrophysical Observatory, Largo Enrico Fermi 5, I-50125 Florence, Italy}
    \and{INAF – Astrophysical Observatory of Catania, Italy}
    \and{MPS, G{\"o}ttingen, Germany}
    \and{INAF – Institute for Space Astrophysics and Cosmic Physics, Milan, Italy}
    }     
   \date{}

 
  \abstract
   {Galactic cosmic rays (GCRs) and solar particles with energies greater than tens of MeV  penetrate spacecraft and instruments hosted aboard space missions.  The Solar Orbiter Metis coronagraph is aimed  at observing the solar corona in both visible  (VL) and ultraviolet (UV) light. Particle tracks are observed in the Metis images of the corona.  
   An algorithm has been implemented in the Metis processing electronics to detect the VL image pixels crossed by cosmic rays. 
   This algorithm  was initially enabled for the VL instrument only,  since
   the process of separating the particle tracks in the UV images has proven to be very challenging. }
   {We study  the impact of the overall bulk of particles of galactic and solar origin on  the Metis coronagraph images. We discuss the effects of the increasing solar activity after the Solar Orbiter mission launch on the secondary particle production in the spacecraft.}
   {We compared Monte Carlo simulations of GCRs crossing or interacting in the Metis VL CMOS sensor to observations gathered in 2020 and 2022. We also evaluated the impact of solar energetic particle events of different intensities on the Metis images.}
   {
  The study of the role of abundant and rare cosmic rays in firing pixels in the  Metis VL  images of the corona allows us to estimate the efficiency of the algorithm applied for cosmic-ray track removal from the images and to demonstrate that the instrument performance had remained unchanged during the first two years of the Solar Orbiter operations.
 The outcome of this work can be used to estimate the Solar Orbiter instrument's deep charging and the order of magnitude for energetic particles crossing the images of Metis and other instruments such as STIX and EUI.
}
   {}
  \maketitle
%

\section{\label{intro}Introduction}
The ESA/NASA Solar Orbiter spacecraft \citep[S/C,][]{a&asoloscience,a&asolomission} hosts six remote sensing and four in situ instruments to image the Sun and to monitor the  plasma and the interplanetary magnetic field, respectively. This mission was launched on  February 10, 2020 at 4:03 UT from Cape Canaveral (Florida, USA).
The S/C cruise period has been characterized by an epoch of minimum-to-low solar activity during the increasing phase of  solar cycle 25   and a positive polarity period of the global solar magnetic field (GSMF). The next GSMF polarity change is expected at the maximum of solar cycle 25 between 2024 and 2025 \citep{singh19}. 
The Solar Orbiter S/C will orbit the Sun between  0.28 AU and 1 AU, with a maximum inclination about the solar equator of 33 degrees during the mission operations. 

Metis is the Solar Orbiter coronagraph aimed at imaging the solar corona in  visible (VL, in the range 580-640 nm) and ultraviolet light \citep[UV, in a $\simeq$ 20 nm band  around the 121.6 nm Lyman-$\alpha$ line,][]{a&ametis,a&ametis2}. The Metis instrument is credited with the first direct imaging of the plasma counterpart of a magnetic switchback in the solar corona \citep{tellonisw}.

Galactic cosmic rays (GCRs) and solar energetic particles (SEPs) interact in the Solar Orbiter S/C and instrument materials. 
GCRs consist approximately of 98\% of protons and helium nuclei, 1\% electrons and 1\% nuclei with Z$\geq$3, where percentages are in particle numbers to the total number \citep{papini96}. The GCR flux  variations as a function of radial distance from the Sun, as well as of the latitude outside the solar equator, time, and energy are discussed in \citet{a&aub}. 

Solar particles in the same energy range of GCRs (above tens of MeV/n) mainly consist of protons (approximately 99\% of the total bulk of SEPs, \citealt{reames21}). 

Daily proton and helium data gathered above 450 MeV n$^{-1}$ by the AMS-02 magnetic spectrometer experiment aboard the Space Station were recently published until 2019 \citep{ams22p,ams22he}. Unfortunately, these measurements do not overlap the mission operations of Solar Orbiter and therefore GCR model predictions are adopted as input spectra for
Monte Carlo simulations \citep{flair,flukacern1,flukacern2} aimed at studying the high-energy particle impact on the corona images gathered with both  VL and UV detectors as part of the Metis diagnostics. 

GCR matrices  gathered with the VL instrument in 2022 were visually analyzed for the purposes of carrying out a comparison with a previous work based on 2020 images  and Monte Carlo simulations \citep{a&aub}.
The efficiency 
of the  algorithm 
for the detection of VL image pixels hit by cosmic rays with respect to those fired by photons is estimated by taking into account the role of both abundant and rare cosmic-ray species. 

The increase in spurious pixels fired in the Metis images by solar particles, in addition to GCRs, is discussed here for the first time.
 
In Section \ref{sect2}, we briefly describe the Metis coronagraph. In Section \ref{sect3}, we report the energy spectra of galactic nuclei, electrons and positrons  for the first part of the Solar Orbiter mission in the summer 2020. The estimates of the proton and helium energy spectra  for the year 2022 are also presented. The role of rare GCR particles in 2022 is inferred from the 2020 analysis.
 In Section \ref{sect4}, we report the energy spectra of protons observed during solar energetic particle events of different intensities. In Section \ref{newana}, we discuss the results of a visual analysis of cosmic-ray matrices gathered in 2020 and 2022. Finally, in Section \ref{sect6}, the Monte Carlo estimates of the number of pixels fired by abundant and rare cosmic rays in the Metis VL and UV images are compared  to cosmic-ray matrix observations gathered with the VL instrument. In addition, the role of SEP events is evaluated.
 
 \section{\label{sect2}The Metis coronagraph}
Metis is the Solar Orbiter coronagraph aimed at carrying out the first simultaneous imaging of the off-limb solar corona in both VL in the range 580-640 nm and 
UV H$\textsc{i}$ Lyman-$\alpha$ line at 121.6 nm. The instrument and first light images are described in detail in \citet{a&ametis,fineschi2020, a&ametis2}  and \citet{antonucci23}. 

The Metis design was optimized to achieve a sensitivity to observe the weak corona from 1.7 R$_{\odot}$ through 9 R$_{\odot}$ by maintaining a contrast ratio lower than 10$^{-9}$  pointing the Sun center within one arcmin. 
The VL detector consists of a VL camera with an active silicon CMOS (CMOSIS ISPHI Rev. B developed by  CMOSIS Imaging Sensors, now AMS-OSRAM\footnote{https://ams-osram.com}, Belgium) sensor segmented in 4.1943 $\times$ 10$^6$ pixels. Each pixel has dimensions of 
10 $\mu$m $\times$ 10 $\mu$m $\times$ 4.5 $\mu$m \citep{a&ametis}. The  geometrical factor of each pixel is   401 $\mu$m$^2$ sr  
\citep{sullivan}. 

The detailed design of the UV detector is illustrated in \citet{uslenghi2017} and \citet{udo18}. It consists of a microchannel
plate (MCP) enclosed in vacuum by 
 a magnesium fluoride entrance window 4 mm thick and a fiber-optic output coupler. 
The MCP has a photocathode coating of potassium bromide (KBr). The UV
radiation  is converted into electrons that are accelerated against a phosphorus
screen. The visible radiation emitted by the screen is captured by
a camera system with a STAR1000 
image sensor of 1024 $\times$ 1024 pixels (for a total of 1.048576 $\times$ 10$^6$ pixels) consisting of two units of
15 $\mu$m $\times$ 15 $\mu$m $\times$ 5 $\mu$m dimensions.
The geometrical factor of each  unit is 743 $\mu$m$^2$ sr.

\section{\label{sect3}Galactic cosmic-ray energy spectra after the Solar Orbiter launch}

 The comparison of the PAMELA and Ulysses experiments' proton data in the energy range 0.92-1.09 GeV has shown  that  the GCR intensity changes by +2.7\% AU$^{-1}$ with  increasing  radial distance from the Sun while a negative variation of 0.024$\pm$ 0.005\% degree$^{-1}$ is observed with increasing heliolatitude  \citep{desimone}. These findings are in agreement, within errors, with those inferred by \citet{helios6}, demonstrating that GCRs show radial gradients of 6.6$\pm$4\% above 50 MeV and 2$\pm$2.5\% in the energy range 250-700 MeV between 0.4 and 1 AU. 
It can be concluded that GCR flux predictions for Solar Orbiter can be carried out with models optimized with data gathered near Earth.

The Gleeson and Axford model \citep[G\&A;][]{glax68} is used  to estimate the  GCR energy spectra  for the Metis diagnostics \citep[see also][]{solwind14, a&aub}. This model  allows us to predict  the cosmic-ray intensity in the inner heliosphere by considering an interstellar energy spectrum and a solar modulation parameter ($\phi$)
used to account for the energy loss of cosmic rays propagating from  the interstellar medium
to  the point of observations in the inner heliosphere. During GSMF positive polarity epochs, the G\&A model has been found to aptly reproduce the  GCR measurements at 1 AU in the energy 
range  from tens of MeV to hundreds of GeV \citep{grim07}. 

\begin{figure}[ht]
\begin{center}
\centering \includegraphics[width=\hsize]{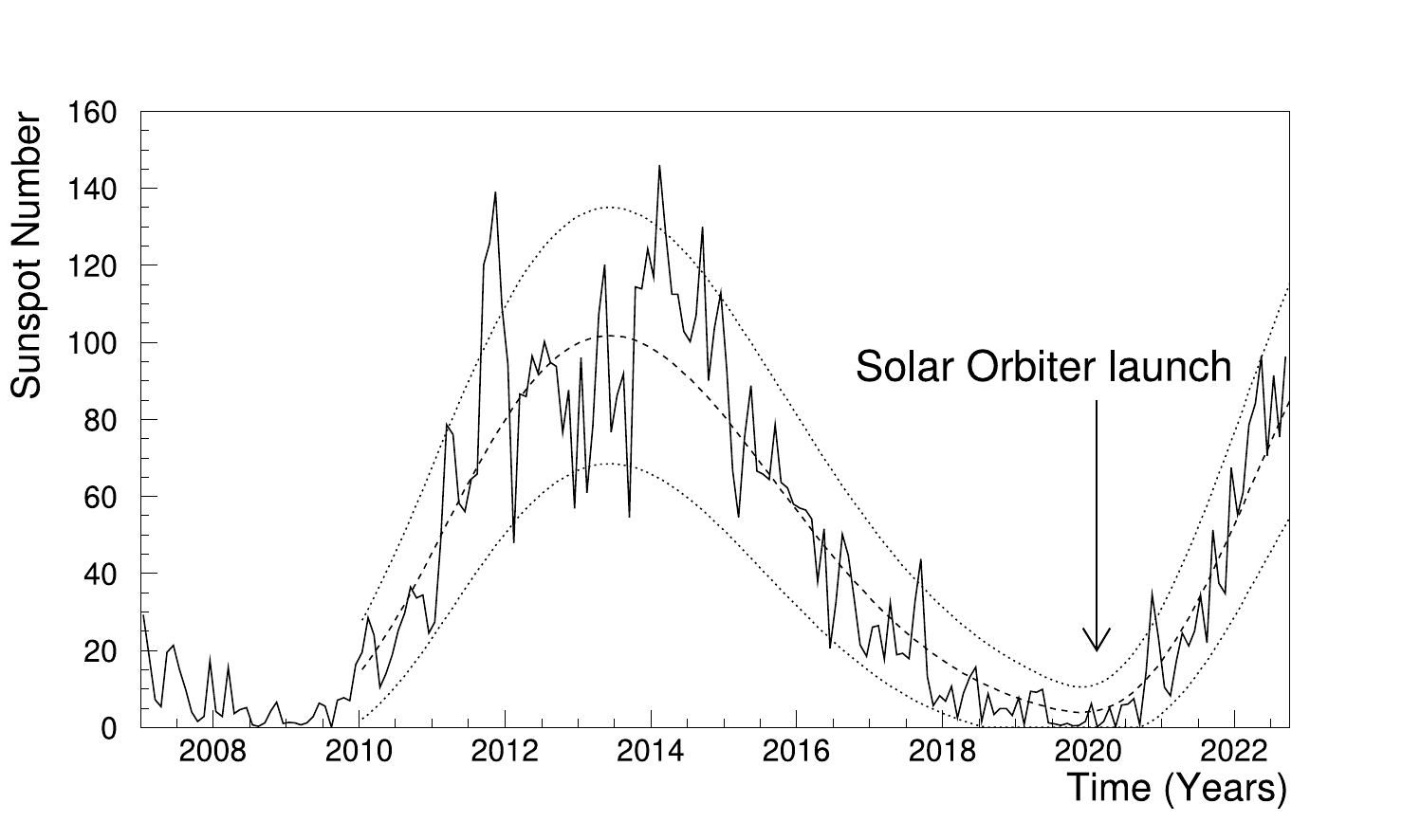}
\end{center}
\caption{\label{ssn} Average monthly sunspot number observed since 2006 during solar cycles 24 and 25. Minimum and maximum predictions  (top and bottom dotted lines) are gathered from \url{http://solarcyclescience.com/solarcycle.html}.}
\end{figure}

The correlation between  the solar modulation parameter\footnote{See also \url{http://cosmicrays.oulu.fi/phi/Phi_mon.txt}} and the solar activity is discussed, for instance, in \citet{brehm21}. The sunspot number, the most widely used proxy for solar activity,  is reported in Fig.~\ref{ssn}\footnote{Data used here are publicly available at \url{http://www.sidc.be/silso/datafiles}} for solar cycle 24 and the first part of solar cycle 25  \citep[see][for details about sunspot number calibration]{silso}.

It must be stressed that the solar modulation parameter assumes different values for similar conditions of solar activity in case different GCR energy spectra  at the interstellar medium are adopted. 
To this end, we used the solar modulation parameter reported in  \citet{uso11, uso17}  estimated with the  \citet{burger2000} interstellar proton spectrum.   Unfortunately, no $^4$He interstellar spectrum is reported in \citet{burger2000}. We choose to use the $^4$He interstellar spectrum inferred from the balloon-borne BESS experiment data \citep{Shikaze2007154, bess14} that were  gathered during different solar modulation and solar polarity periods.  

We took the opportunity to verify the reliability of the approach we use for GCR predictions with the LISA Pathfinder mission for which we estimated the  charging of the interferometer mirrors in 2016 during solar modulation conditions very similar to those of 2022 \citep[see][for details]{bridge}. Proton and helium model predictions were compared to monthly averaged observations of the  AMS-02 experiment available until 2019. An excellent  agreement was found for protons \citep{ams22p}, whereas the model presented a 20\% excess for $^4$He with respect to measurements  \citep{ams22he}. 
The effect of depressing the $^4$He flux model predictions by 20\% on the simulations is discussed in Section 6.

\begin{figure}[ht]
\begin{center}
\centering \includegraphics[width=\hsize]{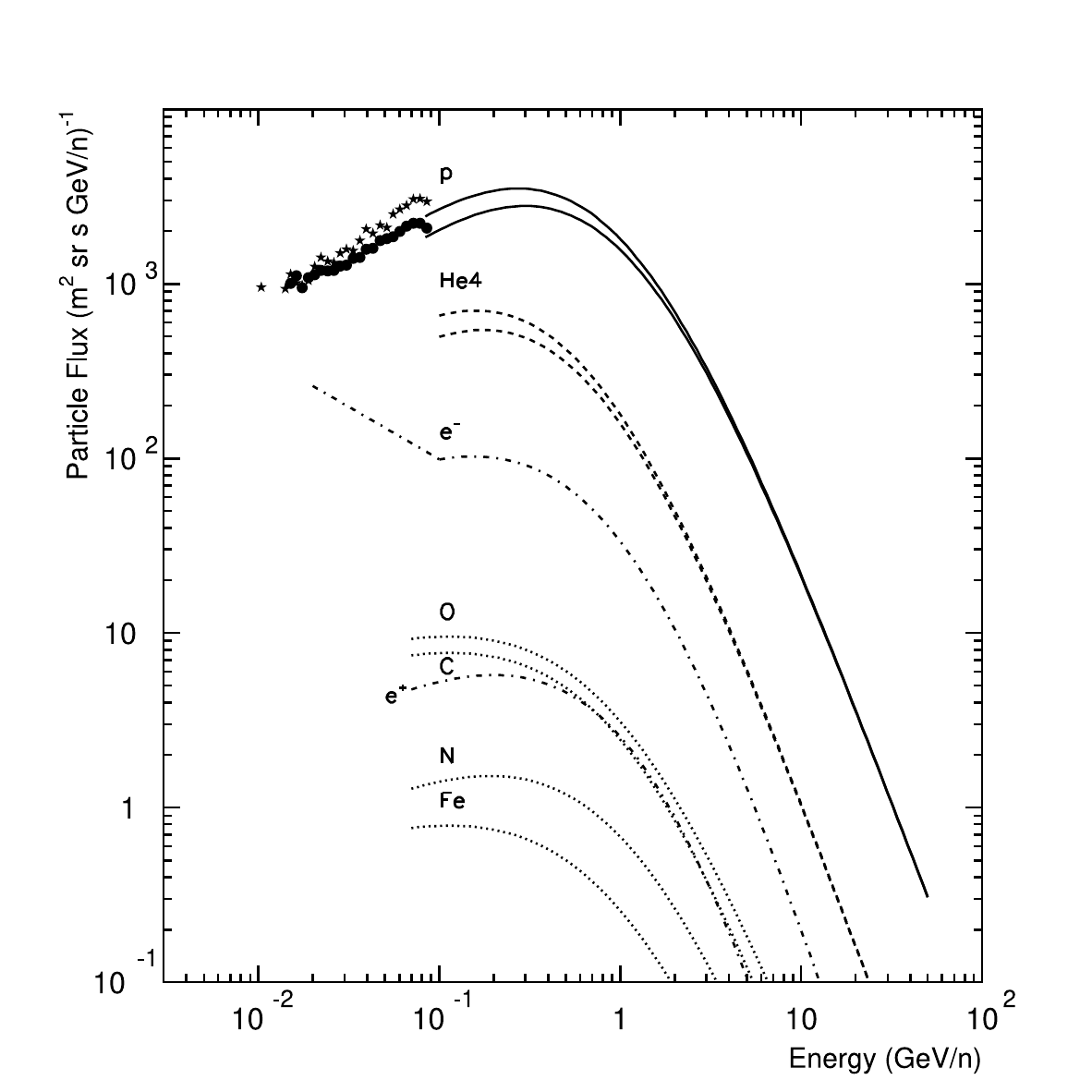}
\end{center}
\caption{\label{metpred} Predictions of cosmic-ray energy spectra after  the Solar Orbiter launch. The top and bottom continuous (protons) and dashed ($^4$He nuclei) curves are obtained  with the G\&A model above 70 MeV(/n) for minimum and maximum solar modulation parameter values of 300 MV/c and 340 MV/c, respectively.  Maximum ($\phi$=300 MV/c) and minimum  ($\phi$=340 MV/c)  proton and helium  predictions apply  to years 2020 and 2022, respectively. The dot-dashed (dotted) curves indicate the electron and positron (nucleus) energy spectra at solar minimum. 
}
\end{figure}

In Fig.~\ref{metpred}, we compare our cosmic-ray predictions to
proton  differential flux measurements gathered in 2020 (solid stars) and in 2022 (solid dots) by the EPD/HET instrument hosted aboard  Solar Orbiter. The  EPD/HET data are publicly available on the Solar Orbiter Archive (SOAR)\footnote{\url{https://soar.esac.esa.int/soar/}}.

In \cite{a&aub}, we demonstrate that during  conditions of solar modulation similar to those observed in the summer 2020, the minimum and maximum  solar modulation parameter  was plausibly ranging between 300 MV/c and 340 MV/c. This upper limit for $\phi$ was estimated but disregarded because  the proton EPD/HET  observations were compatible with model predictions obtained by assuming $\phi$=300 MV/c within  model uncertainties and measurements \citep{wimmer21}. 

As a matter of fact, the average observed solar modulation parameter in June-July 2020 was 293 MV/c\footnote[2]{\url{http://cosmicrays.oulu.fi/phi/Phi_mon.txt}}. Therefore, for nuclei, electrons, and positrons, in 2020, we considered the near-Earth energy spectra gathered at solar minimum and reported in \citet{papini96}, \citet{gri04}, \citet{gri07}, and \citet{griele}. 

Nucleus, electron, and positron flux model predictions far from solar minimum and maximum conditions are affected by uncertainties larger than the contribution that these particles  give in increasing the number  of  tracks in the Metis images. 
The average solar modulation parameter in 2021 was 327 MV/c and it is plausible to expect for the same not less than 340 MV/c in June-July 2022. On the basis of these considerations and of the EPD/HET proton data shown in Fig.\ref{metpred}, 
 we  adopted the maximum value of the solar modulation parameter set for the Solar Orbiter cruise phase  ($\phi$=340 MV/c) for the year 2022. 

The galactic particle energy spectra above 70 MeV are parameterized as follows (see for details \citealt{apj1}):

\begin{equation}
F(E)= A\ (E+b)^{-\alpha}\ E^{\beta}  \ \ \ {\rm particles/(m^2\ sr\ s\ GeV/n),}
\label{equation1}
\end{equation}

\noindent where $E$ is the particle kinetic energy in  GeV/n.
The parameters $A$, $b$, $\alpha$, and $\beta$  for  solar minimum in the summer 2020  are reported in Table~\ref{table1}. The units of the parameters $A$ and $b$ are particles/(m$^2$ sr s (GeV/n)$^{-\alpha +\beta +1}$) and GeV/n, respectively, while the spectral indices $\alpha$ and $\beta$ are pure numbers.

 A power-law interpolation function was used    below 100 MeV for electron energy spectra, and above 20 GeV for positrons \citep{griele}:

\begin{equation}
F(E)= A\ E^{-\beta}  \ \ \ {\rm particles/(m^2\ sr\ s\ GeV).}
\label{equation2}
\end{equation}

\noindent In this last  equation, $A$ is measured in  particles/(m$^2$ sr s GeV$^{-\beta+1}$) and $\beta$ is a pure number. 

 In Table~\ref{table1} (from top to bottom), the energy ranges of  the parameterizations of the electron flux  are:  0.02 GeV-0.1 GeV, and 0.1 GeV-200 GeV. For positrons, they are: 0.07 GeV-20 GeV, and 20 GeV-200 GeV.
 Proton and helium energy spectra parameterizations for the year 2022 are reported in Table~\ref{table2}. All particle energy spectra are shown in Fig.~\ref{metpred}. 
 
 As it was mentioned above, due to lack of continuous data gathering in space for electrons, positrons and heavy nuclei,  models cannot be tested against observations carried out during intermediate solar activity conditions.  Predictions of rare cosmic-ray particle energy spectra  were not considered for 2022 due to  uncertainties that could potentially be introduced by the model. The contribution of these particles will be estimated  in Section 6 on the basis of the results obtained at solar minimum.

\begin{table}  
\caption{\label{table1} Parameterizations of cosmic-ray energy spectra 
 in June-July 2020.  The parameterizations of the energy spectra of protons and nuclei are meant above 70 MeV(/n).}    
\centering
\begin{tabular}{@{}*{5}{l}}
\hline  
\hline
Particle species&  $A$ &  $b$ &  $\alpha$ &  $\beta$ \\ 
\hline 
p & 18000. & 0.875 & 3.66 & 0.87 \\                      
He & 850. & 0.53 & 3.68 & 0.85\\
C & 23. & 0.95 & 3.00 & 0.32\\
O& 25.2 & 1.05 & 3.25 & 0.32\\
N& 7.0 & 1.05 & 3.25 & 0.5\\
Fe& 1.9 & 0.95 & 3.00 & 0.32\\
e$^-$ E $\leq$ 0.1 GeV& 24.7& & &0.60\\
e$^-$ E $>$ 0.1 GeV & 400.& 0.97& 3.66&0.5\\
e$^+$ E $\leq$ 20 GeV& 100.& 1.45&4.1&0.5\\
e$^+$ E $>$ 20 GeV& 7.61& & &2.84\\
\hline
\end{tabular}
\end{table}  

\begin{table}  
\caption{\label{table2}  
Same as Table~\ref{table1} for protons and helium energy spectra in May 2022.}    
\centering
\begin{tabular}{@{}*{5}{l}}
\hline  
\hline
Particle species &  $A$ &  $b$ &  $\alpha$ &  $\beta$ \\ 
\hline 
p & 18000. & 0.95 & 3.66 & 0.87 \\ 
He & 850. & 0.58 & 3.68 & 0.85\\
\hline
\end{tabular}
\end{table}

\section{\label{sect4}Solar energetic particles}
The Sun flings one million tons of fully ionized plasma out from the corona every second. The expanding solar wind drags the solar magnetic field forming the interplanetary magnetic field. The typical energy of the solar wind particles is of 0.5-3 keV. Particles of solar origin with energies larger than 1 MeV are observed during "impulsive" and "gradual" events \citep{reames21}. Short-duration ($\simeq$ hours) impulsive events are generated by magnetic reconnection on open field lines in solar jets, while  long-duration ($\simeq$ days) gradual events are associated with coronal mass ejections driving shock waves. Pure impulsive or gradual events are rare and shock waves may
reaccelerate suprathermal particles from impulsive events. 
Impulsive and gradual events present different particle energy spectra. Particle acceleration is limited to about 50 MeV during impulsive events. Consequently, due to the average grammage of several g cm$^{-2}$ of S/C and instrument materials stopping low-energy particles before reaching the sensitive parts of the Metis instrument, we estimate the number of pixels fired  in the VL and UV images of the solar corona during gradual events only. 

It is worthwhile to point out that above tens of MeV,  protons overcome by approximately two orders of magnitude the other  species of particles during gradual SEP events, as observed in space by the PAMELA magnetic spectrometer experiment that  monitored  both proton and helium differential fluxes during the evolution of two gradual SEP events dated December 13 and December 14, 2006 \citep{pamFD} up to GeV energies. These two SEP events were characterized by fluences ranging between 10$^5$ and  10$^7$ protons cm$^{-2}$ above 70 MeV. The onset of the December 13, 2006 event was observed between 03:18 UT and 03:45 UT, while data were gathered at the peak between 04:33 UT and 04:59 UT. The whole event duration was of two days. The onset of the weak December 14, 2006 event was observed between 23:05 UT on December 14, 2006 and 02:35 UT on December 15, 2006, while the decay phase was measured between 19:30 UT and 23:35 UT of the second day. 

\begin{figure}[ht]
\begin{center}
\centering \includegraphics[width=\hsize]{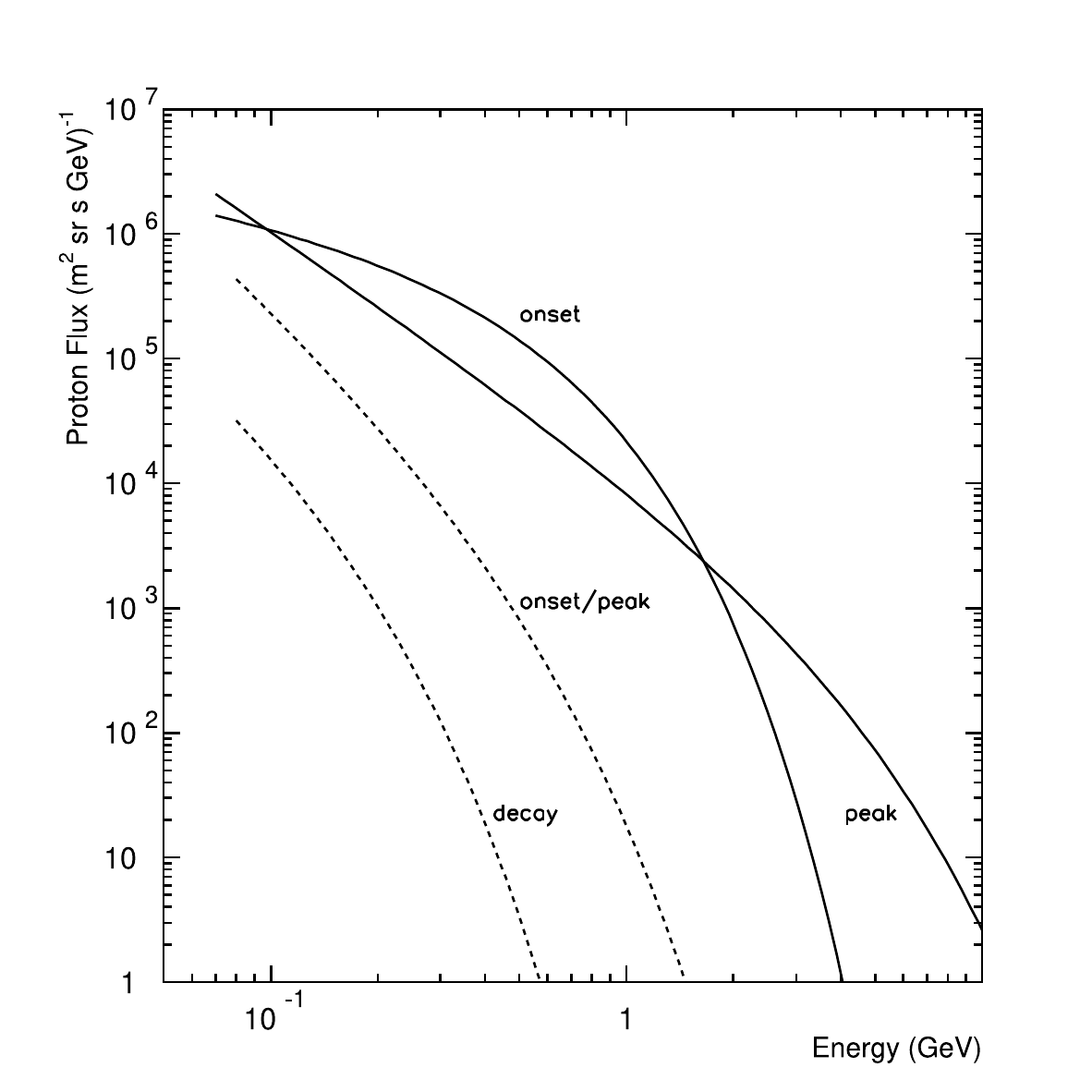}
\end{center}
\caption{\label{pamela} Solar proton energy spectra measured by the PAMELA experiment during the evolution of the SEP events dated December 13 (solid lines) and December 14 (dashed lines), 2006. The different phases of the events are indicated in the figure.
}
\end{figure}

We focus on these  SEP events because data gathered in space with a magnetic spectrometer are certainly the most accurate at high energies to be adopted for Monte Carlo simulations.
Unfortunately, most space experiments  devoted to solar particle monitoring do not allow for measurements of the particle differential fluxes above 100 MeV(/n). This may most likely be ascribed to evidence that the most frequent solar particle events are characterized by particle acceleration well below hundreds of MeV, even though the major space weather events that lead to substantial S/C inner charging are associated with particles of GeV energies. These SEP events exceed the GCR background by several orders of magnitude. The proton energy spectra observed during the evolution of the December 13 and December 14, 2006  SEP events are shown in Fig.~\ref{pamela} \cite[see also][]{spectral}.

\begin{figure}[h!]
\begin{center}
\centering \includegraphics[width=0.9\hsize]{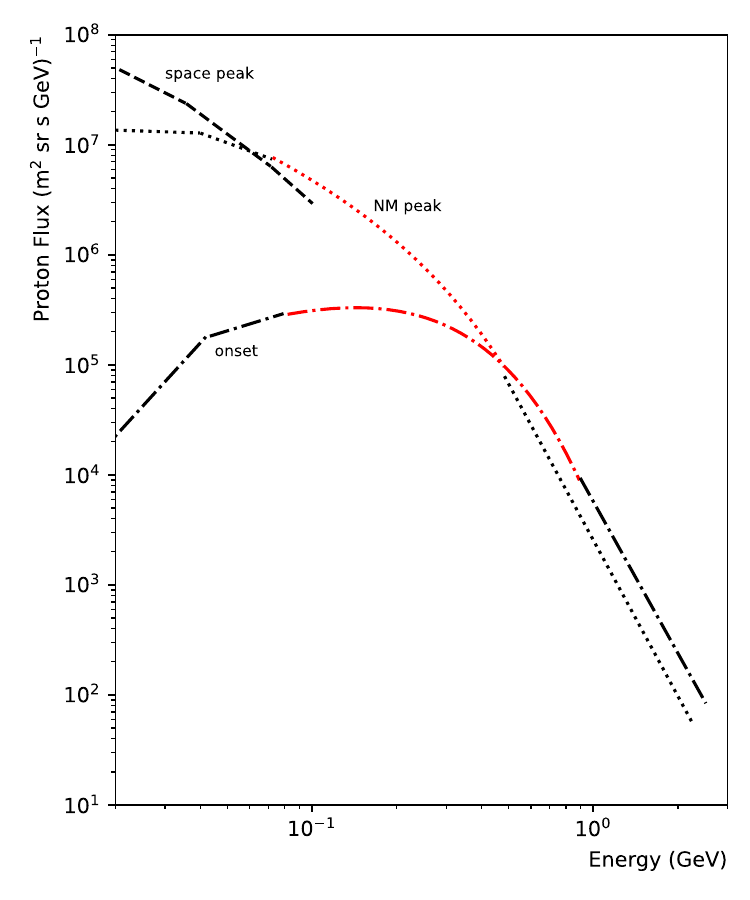}
\end{center}
\caption{\label{prot28} Solar proton energy spectra observed during the evolution of the solar energetic particle event dated October 28, 2021. The dot-dashed line indicates the onset of the event (15:35-16:35 UT). The dotted line corresponds to the peak of the event observed on ground (17:30-18:20 UT) and the dashed line represents the peak of the event in space (20:35-22:35 UT). Low-energy EPD/HET data below 100 MeV and  NM observations above 700 MeV have been interpolated between 100 MeV and 700 MeV according to \citet{spectral} (red lines). 
}
\end{figure}

A more recent event characterized by particle acceleration above 2 GeV was  observed   
on October 28, 2021  by several instruments in space, including EPD/HET aboard Solar Orbiter, and with neutron monitors (NMs) on Earth. During  this event, Solar Orbiter was almost radially aligned with Earth \citep[see for details][]{papa22,martucci23}. The solar eruption started with an X1.0 class flare at 15:17 UT and peaked at 15:35 UT. This event generated the first ground level enhancement for solar cycle 25.
The onset of the event was detected aboard Solar Orbiter  at 15:35 UT by EPD/HET. 
Neutron monitor data, binned every 30 minutes, reported the onset at 16:00 UT consistently with the Solar Orbiter measurements. The observed solar proton flux at the onset is represented by the dot-dashed curve in Fig.~\ref{prot28}. The contemporaneous availability of the low-energy space data below 100 MeV and observations above 700 MeV gathered on Earth, allowed us to interpolate the two data sets in the 100-700 MeV range. 

The most energetic particles are observed at the onset of SEP events, while lower energy particles appear in increasing number at the peak of the event when the high-energy particles fade away \citep{dalla03}. 
In particular, when particle acceleration  occurs below 500 MeV, the atmosphere shielding prevents secondary particle production to enhance the NM counting rate,  while a large flux of low-energy particles may be associated with the peak of the event in space. As a matter of fact, the October 28, 2021 event peak  was observed on Earth at 18:00 UT (dotted curve in Fig.~\ref{prot28}), while the peak in space was detected between 20:35 UT and 22:35 UT  (dashed curve in Fig.~\ref{prot28}). No enhancement of the NM counting rate was observed on Earth at this time. As a result, the peak proton flux in space was not interpolated at  energies above the range of availability of the Solar Orbiter/HET data.

Unfortunately, no Metis VL cosmic-ray matrices were acquired during the period of the October 28, 2021 SEP event. On the other hand, the UV images do not allow us to carry out an effective visual analysis of particle tracks  due to the high number of spurious fired pixels present in the images resulting very difficult to separate from genuine photon signals. Here, we considered Monte Carlo simulations to estimate the number of pixels fired by solar protons in the Metis images  during SEP events as those described above.

\section{\label{newana} Visual analysis of VL cosmic-ray matrices in 2022}
In \citet{a&aub} we have reported the outcomes of a visual analysis of cosmic-ray tracks in 
four sets of four co-added 15-second cosmic-ray matrices, for a total exposure time of 60 seconds. These images were taken on May 29, 2020. 

An algorithm, described in detail in the above work, allows us to separate pixels fired by VL photons from those crossed  by high-energy particles depositing a larger amount of energy by ionization in the CMOS of the VL instrument  \citep[see also][]{andre}. An average of 271$\pm$22 cosmic-ray tracks per set of images were observed  after removing the noisy pixels found in more than one image of each set, corresponding to a fraction of about 10$^{-5}$ of the total number of image pixel sample. The noisy pixels were found one order of magnitude smaller with respect to those fired by cosmic rays. The lower limit to the cosmic-ray pixel firing  efficiency was estimated equal to 0.94$\pm$0.02 by studying a sample of particle slant tracks.

A new analysis  of images gathered from May 9 through May 15, 2022 was carried out to test the stability of the instrument performance 
after two years during the increasing phase of the solar cycle 25. We studied three sets of  14 superposed frames of 30 seconds each comprise a total 7 minute exposure time. 
The visual analysis was carried out with the APViewer adapted to the 
new sets of images and described in detail in \cite{a&aub}.

Noisy single pixels, clusters and columns of fired pixels appearing in more than two images have been removed. The percentage of spurious pixels and the efficiency of single pixels have been found compatible with the first analysis. This evidence indicates  that the performance of the VL instrument remained unchanged during the first two years of the Solar Orbiter mission.

\begin{figure*}[h!]
\begin{center}
\centering \includegraphics[width=0.7\hsize]{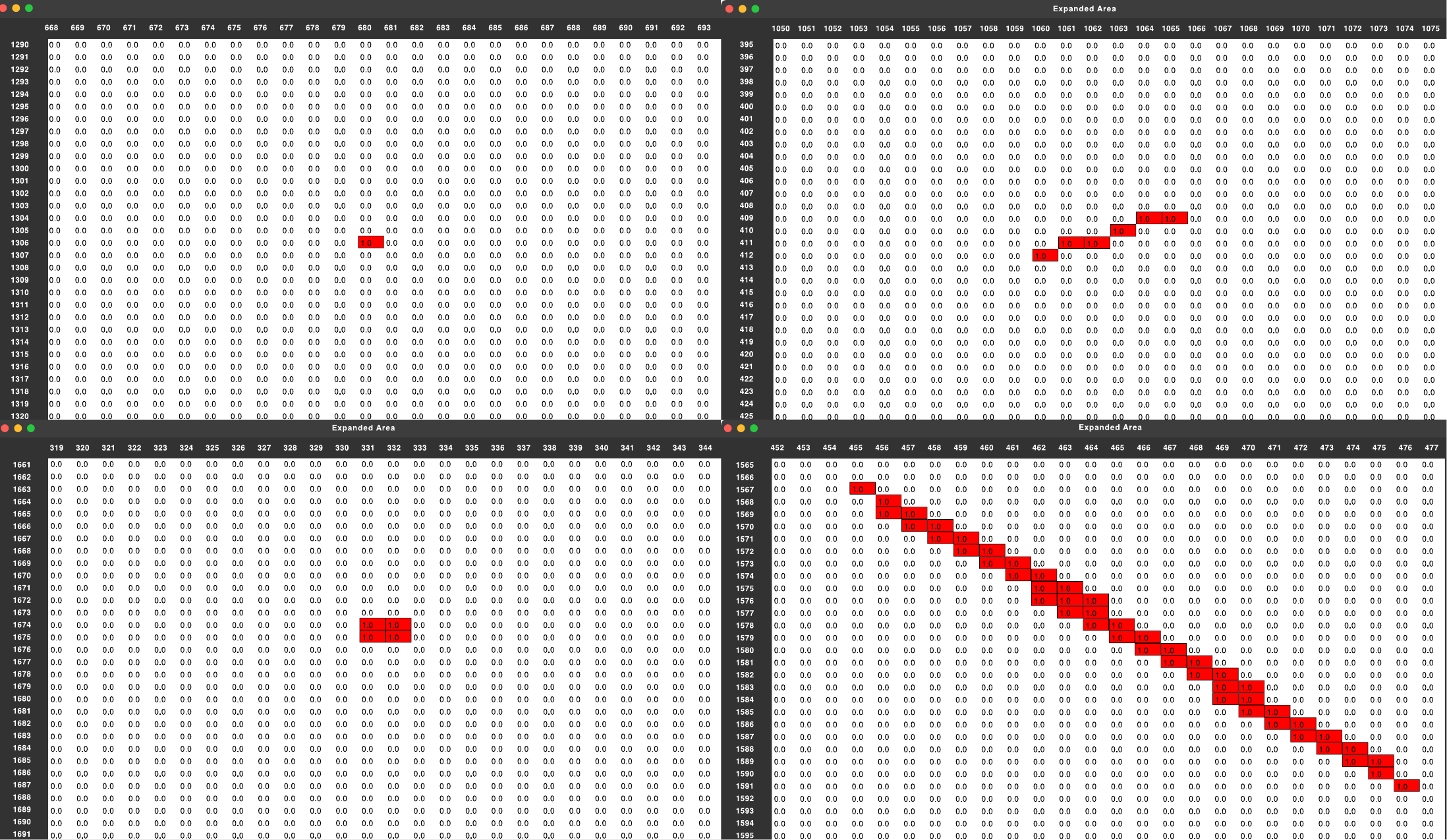}
\end{center}
\caption{\label{persici} Particle tracks in the Metis VL cosmic-ray matrices gathered in 2022. Top-left panel represents a single fired pixel associated with straight tracks, while the top-right panel shows a typical slant track. The bottom left and right panels report a square and an exceptional composite track, respectively. Both squares and composite tracks are considered to be formed by  pixels  in the same line crossed by the incident particle and extra pixels fired by photons and knock-on electrons generated along the main cosmic-ray track.  Future works will allow us to verify this hypothesis.  
}
\end{figure*}

Particle straight tracks (single fired pixels) and slant tracks were easily identified. In addition, we found samples of  clusters of pixels 
compatible with a main particle track and side fired pixels.  
These clusters are displayed as "squares" and "composite tracks" in Fig. \ref{persici}. 

The total number of straight tracks, slant tracks firing more  than one pixel, with or without  extra pixels fired along the main particle track, are reported in Table~\ref{table3} after the normalization of data to a 60 second exposure time, meant for a comparison with the first analysis. The average number of particle tracks of 212$\pm$6 is observed  to be smaller with respect to the 2020 observations. This is expected due to the increasing solar activity during the last two years  and the consequent reduction of the GCR flux. 

\begin{table}[h!]  
\caption{\label{table3} Metis cosmic-ray observations in three 7-minute exposure time images gathered in May 2022 by the Metis VL instrument. Data have been normalized to one minute exposure time for comparison with the first analysis carried out at solar minimum in 2020. Examples of the track topology are reported in Fig. \ref{persici}.}
\centering
\begin{tabular}{@{}*{6}{l}}
\hline  
\hline
 & Straight & Slant & Squares & Total & Composite \\ 
\hline 
May 2022&&&&& \\
Image 1 & 165 & 60 & 2 & 227 &  19\\         
Image 2  &  130 & 63 & 6  & 199 &  23\\
Image 3  &  159 & 49 & 3 & 211&  19\\
\hline
 Average  &151 & 57 & 4& 212 $\pm$ 6 & 20\\
\hline
May 2020&&&&& \\
Average & 188 & 79 & 4 & 271$\pm$22 &23  \\
\hline
\end{tabular}
\end{table}


\section{\label{sect6}Monte Carlo simulations of  high-energy particles firing spurious pixels in the Metis VL and UV images}
\subsection{Galactic cosmic rays}
\begin{figure}[ht]
\begin{center}
\centering \includegraphics[width=0.9\hsize]{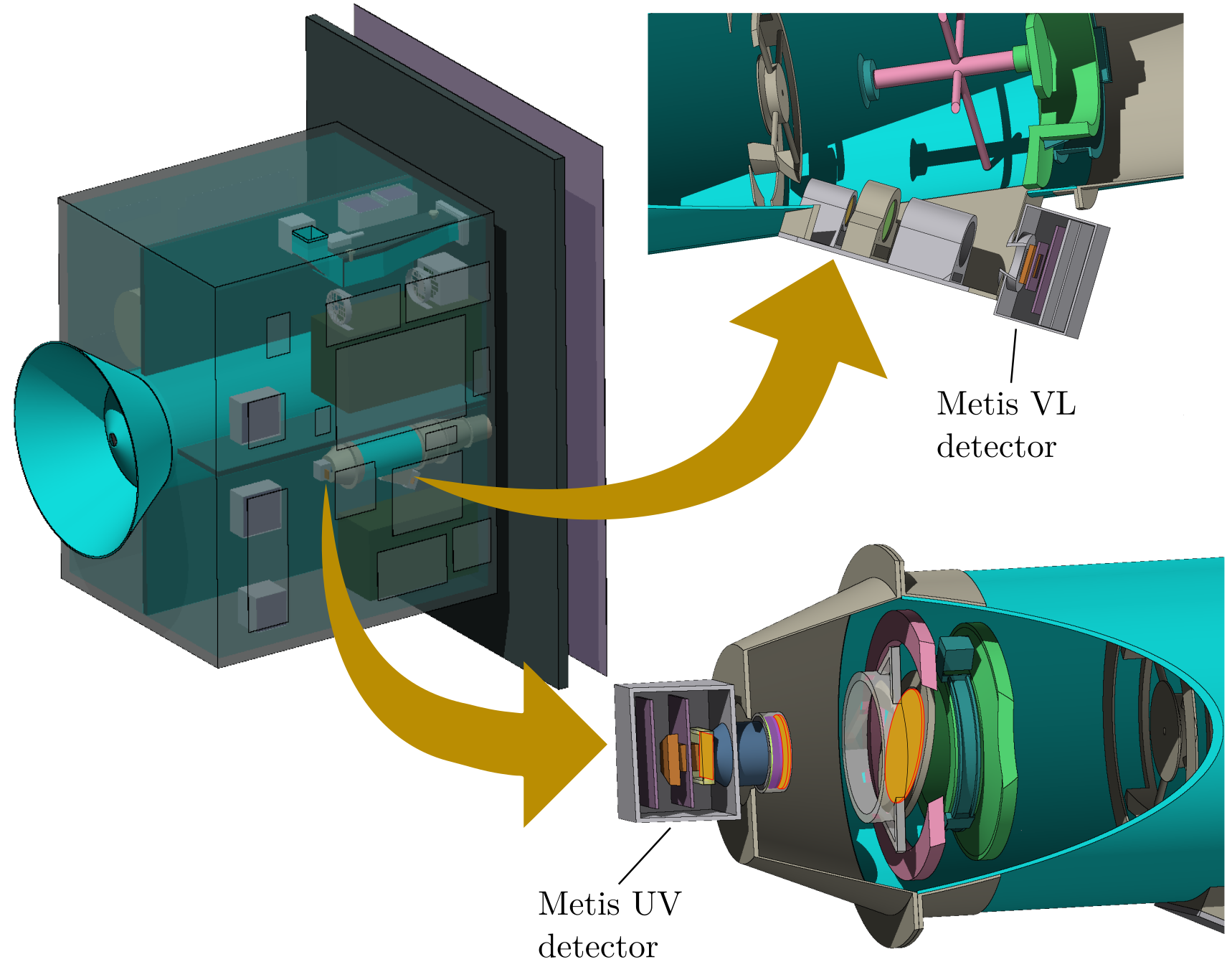}
\end{center}
\caption{\label{flukamodel} Solar Orbiter geometrical model. Remote sensing instruments and electronic boxes are visible. The  magnified images show the VL and UV detectors.
}
\end{figure}
The geometry of the Solar Orbiter S/C and instruments built with Flair \citep{flair} in  FLUKA (version 4.0.1) for the Metis VL and UV detector simulations is shown in Fig.~\ref{flukamodel}. Details of the VL and UV instruments are shown in the magnified images. The geometry includes the S/C structure, thrusters, fuel tanks and the SPICE, EUI, PHI, and STIX instruments \citep[][and references therein]{a&asolomission}. Interplanetary and galactic particles cross from about 1 g cm$^{-2}$ to more than 10 g cm$^{-2}$ of material depending on the particle incidence direction before reaching the Metis VL and UV instruments.

The simulations for the year 2020 returned 276$\pm$17  pixels fired by incident protons only in 60 seconds of exposure time.
This number of tracks appeared similar to the observations within the statistical uncertainties. 

It was suggested that the Metis VL detector could play the role of a proton monitor, in cases where the efficiency of the algorithm for cosmic-ray track removal from VL images was approximately  35\% of the proton contribution; that is, numerically equivalent to the sample of particle tracks generated  by the other components of GCRs \citep{a&aub}.  The actual efficiency of the VL instrument algorithm  for cosmic-ray detection is   estimated here since the solar modulation parameter in the summer 2020 is now known. 

The  visual analysis of the Metis VL images does not allow us to disentangle primary and secondary particles and the different kinds of particles. As a result, the Monte Carlo simulations play a primary role to study  the number of single and clustered pixels fired by high-energy particles  in the Metis cosmic-ray matrices.
 Rare particle energy spectra are determined at solar minimum 
  according to \citet{papini96} and \citet{bridge}. 
During periods of low solar activity simulation outcomes are affected by two systematic uncertainties of about 10\% associated with cosmic-ray models and  FLUKA Monte Carlo program accuracy \citep{Lechner19}.

In the majority of cases, cosmic rays fire single pixels in the Metis images. Observations and simulations return a similar number of slant-to-straight  tracks compatible with the isotropic distribution of cosmic rays incident on the Solar Orbiter spacecraft and the geometrical shape of the sensitive parts of the instruments.

The contributions of nuclei, electrons and positrons  to the overall sample of pixels fired by cosmic rays  reported in Table~\ref{table4} actually amount to 38\%, thus confirming our prediction of the VL instrument algorithm efficiency for cosmic-ray selection. Simulations reveal also that cosmic rays fire approximately twice the number of pixels in the UV images  with respect to those in the VL images mainly because of a larger  geometrical factor of the UV instrument, but also because of a different amount of matter  surrounding the sensitive parts of the UV and VL  cameras.  
Due to the increasing solar activity  over the last two years, in 2022 the proton and helium fluxes are expected to show a decrease of no less than 15\% with respect to  summer 2020. 

The  simulated number of spurious pixels fired in the VL images by cosmic-ray protons and helium nuclei decreases by 13\%  as can be observed in Table~\ref{table5} (numbers in parentheses for $^4$He). With respect to the helium nuclei contribution, Monte Carlo simulations indicate that if a reduction of the input flux of 20\% is considered, the  contribution of these nuclei to the overall sample of tracks is decreased by 3\%. This effect was not considered in the 2020 analysis, but it is taken into account in the 2022 simulations. 

An additional plausible contribution of rare particles of  10\% of the sample of tracks generated by protons and $^4$He should be added to these estimates  on the basis of the 2020 results and the assumption of a slightly higher solar modulation parameter in 2022 with respect to 2020.  

The analysis of the 2022 VL cosmic-ray matrices returns a smaller number of observed tracks  with respect to simulations. Systematic and statistical uncertainties on simulation results are reported in  Tables~\ref{table4} and \ref{table5}. Simulations and observations are in agreement within slightly more than two standard deviations. We will study the role of the actual solar modulation parameter in summer 2022 in future works.

A smaller decrease in the proportion of fired pixels in the UV images (9\%) with respect to VL images was found with the simulations.
The summer 2020 simulations indicated that for primary protons, the charged particles crossing the VL images in particle numbers to the total number consisted of 80\% protons, 17\% electrons and positrons, and 3\% pions. For the year 2022, protons are estimated to fire 77\% (72\%) of the total sample of spurious pixels in the VL (UV) images, while 18\% (22\%) of pixels are hit by electrons and positrons and then 5\% (6\%) by pions. 
 The ratio of secondary-to-primary particles in the UV images appears 5\% larger than in the VL images. At the next solar maximum  we will test the capability of the VL instrument to work as a proton monitor in comparison to the UV instrument.
This work allows us to study the composition of high-energy particles deep into the Solar Orbiter S/C  and may result of interest to other instruments such as EUI and STIX.

\subsection{Solar energetic particle events}
During gradual SEP events, the overall flux of particles observed in space increases by several orders of magnitude. The majority of these events present a fluence in the range 10$^6$-10$^7$ protons cm$^{-2}$ above 30 MeV \citep{nym99a,nym99b}. The events described  in Section \ref{sect4}  belong to this range of intensities. In Table~\ref{table6} we report the number of pixels expected to be fired by solar particles during the evolution of these events. By comparing these results with those appearing in Table~\ref{table4}, it is possible to notice that during weak-to-medium gradual SEP events, the number of pixels hit by high-energy particles increases by 1-2 orders of magnitude and the fraction of  spurious fired pixels to the total number of pixels varies from 10$^{-4}$ to 10$^{-2}$. Stronger events would further affect the VL and UV instrument observations. These predictions will be verified with VL cosmic-ray matrices gathered in the future during SEP events to assess also the impact of solar particles in the S/C inner charging.

\section{\label{concl}Conclusions}
High-energy particles  interact in the Solar Orbiter S/C, thus limiting the efficiency of on-board instruments.  It is found that for 60 seconds of exposure time near solar minimum, the number of pixels crossed by galactic cosmic rays in the VL images of the Metis coronagraph is  a fraction of about 10$^{-4}$ of the total number of pixels. Monte Carlo simulations of the VL instrument return a similar number of  tracks associated with primary galactic protons. The contribution of cosmic-ray nuclei with charge $>$1, electrons and positrons accounts approximately for the overall efficiency  of the on-board algorithm for high-energy particle detection found of 38\%. 

Simulations of the number of pixels fired in UV images indicate  a larger number of tracks    with respect to those present in the VL images. This is mainly due to a larger geometrical factor  of the UV instrument  and to a different material  distribution around  the two instruments.

The increase of the solar activity during the year 2022 is expected to have reduced the intensity of the GCR flux by  at least  15\% with respect to the summer 2020. Monte Carlo simulations show a similar decrease in pixels fired by GCRs in the VL images, while the number of tracks in the UV images decreases by less than 10\%, after being contaminated by a larger number of secondaries.

The number of observed and simulated tracks in the VL images in 2022 appears in agreement within slightly more than two standard deviations.
The Metis VL instrument has not modified its performance after the mission launch in terms of an excess of spurious fired pixels with respect to the 2020 analysis. 
A smaller number of observed tracks with respect to simulations may indicate that an overly small solar modulation parameter has been assumed for the  simulations.

The Metis VL images enable the monitoring of long-term GCR proton flux variations and SEP event evolution, when compared to Monte Carlo simulations of the instrument performance. 
The number of Metis corona image pixels fired by high-energy particles is expected to increase by 1-2 orders of magnitude during the evolution of medium-strong SEP events. 

This simulation work, meant for the Metis diagnostics, can be also used to study the Solar Orbiter S/C inner charging during the mission operations and to estimate the role of the impact of particle tracks in the images of other instruments such as EUI and STIX. 

\begin{table}  
\caption{\label{table4} Galactic cosmic-ray tracks in the Metis images  from Monte Carlo simulations for a 60 second exposure time in the summer 2020. Systematic,  statistical and total uncertainties are indicated and combined in quadrature.}
\centering
\begin{tabular}{@{}*{3}{l}}
\hline  
\hline
 Particle species & VL detector &  UV detector \\ 
\hline 
Protons & 276 & 442  \\         
Helium & 77 & 110 \\
Carbon & 4 & 7 \\
Nitrogen& 2 & 3\\
Oxygen& 5 & 6\\
Iron& 2 & 2 \\
Electrons& 14 & 24\\
Positrons& 1 & 1 \\
\hline
& 381$\pm53$$\pm20$ & 595$\pm83$$\pm24$ \\
Total & 381$\pm57$ & 595$\pm86$ \\
\end{tabular}
\end{table}

\begin{table}  
\caption{\label{table5} Same as Table~\ref{table4} for the year 2022.  The number of tracks ascribable to $^4$He nuclei in the parentheses indicate the estimate obtained with the \citet{Shikaze2007154} interstellar spectrum before normalization on observed data during a period of similar solar activity.}
\centering
\begin{tabular}{@{}*{3}{l}}
\hline  
\hline
 Particle species & VL detector &  UV detector \\ 
\hline 
Protons & 242 & 402  \\         
Helium & 58 (67)  & 88 (101)  \\
Rare particles & 30 & 49 \\
\hline
 & 330$\pm46$$\pm18$ & 539$\pm75$$\pm23$ \\
Total & 330$\pm49$& 539$\pm78$ \\
\end{tabular}
\end{table}

\begin{table}  
\caption{\label{table6} Monte Carlo simulations of solar energetic particle tracks in the Metis corona images for a 60 second exposure time during typical events of different intensity. 
}    
\centering
\begin{tabular}{@{}*{3}{l}}
\hline  
\hline
 SEP event & VL detector &  UV detector \\ 
\hline 
December 13, 2006 (onset) & 24600 & 56400  \\  
December 13, 2006 (peak)& 11600 & 30800 \\
December 14, 2006 (onset/peak)& 1380 & 1980 \\
December 14, 2006 (decay)& 180 & 201\\
October 28, 2021 (onset) & 12960& 28680\\
October 28, 2021 (NM peak) & 9000 &  13380\\
October 28, 2021 (HET peak) & 38400& 60000 \\
\hline
\end{tabular}
\end{table}

\begin{acknowledgements}
Solar Orbiter is a space mission of international collaboration between ESA and NASA, operated by ESA.  The Metis program is supported by the Italian Space Agency (ASI) under the contracts to the co-financing National Institute of Astrophysics (INAF): Accordi ASI-INAF N. I-043-10-0 and Addendum N. I-013-12-0/1, Accordo ASI-INAF N.2018-30-HH.0  and under the contracts to the industrial partners OHB Italia SpA, Thales Alenia Space Italia SpA and ALTEC: ASI-TASI N. I-037-11-0 and ASI-ATI N. 2013-057-I.0. Metis was built with hardware contributions from Germany (Bundesministerium für Wirtschaft und Energie (BMWi) through the Deutsches Zentrum für Luft- und Raumfahrt e.V. (DLR)), from the Academy of Science of the Czech Republic (PRODEX) and from ESA. \\
We thank J. Pacheco and J. Von Forstner of the EPD/HET collaboration for useful discussions about cosmic-ray data observations gathered aboard Solar Orbiter up to 100 MeV. We also thank the PHI and EUI Collaborations for providing useful details about instrument geometries for S/C simulations.
\end{acknowledgements}

%
   \bibliographystyle{aa} 
   \bibliography{biba25Apr22} 

\begin{thebibliography}{46}
\expandafter\ifx\csname natexlab\endcsname\relax\def\natexlab#1{#1}\fi

\bibitem[{{Abe} {et~al.}(2014){Abe}, {Fuke}, {Haino}, {Hams}, {Hasegawa},
  {Horikoshi}, {Itazaki}, {Kim}, {Kumazawa}, {Kusumoto}, {Lee}, {Makida},
  {Matsuda}, {Matsukawa}, {Matsumoto}, {Mitchell}, {Moiseev}, {Nishimura},
  {Nozaki}, {Orito}, {Ormes}, {Picot-Cl{\'e}mente}, {Sakai}, {Sasaki}, {Seo},
  {Shikaze}, {Shinoda}, {Streitmatter}, {Suzuki}, {Takasugi}, {Takeuchi},
  {Tanaka}, {Thakur}, {Yamagami}, {Yamamoto}, {Yoshida}, \&
  {Yoshimura}}]{bess14}
{Abe}, K., {Fuke}, H., {Haino}, S., {et~al.} 2014, Advances in Space Research,
  53, 1426

\bibitem[{Adriani {et~al.}(2011)Adriani, Barbarino, Bazilevskaya, Bellotti,
  Boezio, Bogomolov, Bonechi, Bongi, Bonvicini, Borisov, Bottai, Bruno,
  Cafagna, Campana, Carbone, Carlson, Casolino, Castellini, Consiglio, Pascale,
  Santis, Simone, Felice, Formato, Galper, Grishantseva, Gillard, Jerse,
  Karelin, Koldashov, Krutkov, Kvashnin, Leonov, Malakhov, Marcelli, Mayorov,
  Menn, Mikhailov, Mocchiutti, Monaco, Mori, Nikonov, Osteria, Palma, Papini,
  Pearce, Picozza, Pizzolotto, Ricci, Ricciarini, Sarkar, Rossetto, Simon,
  Sparvoli, Spillantini, Stozhkov, Vacchi, Vannuccini, Vasilyev, Voronov, Wu,
  Yurkin, Zampa, Zampa, \& Zverev}]{pamFD}
Adriani, O., Barbarino, G.~C., Bazilevskaya, G.~A., {et~al.} 2011, The
  Astrophysical Journal, 742, 102

\bibitem[{Aguilar {et~al.}(2021)Aguilar, Cavasonza, Ambrosi, Arruda, Attig,
  Barao, Barrin, Bartoloni, Ba\ifmmode \mbox{\c{s}}\else \c{s}\fi{}e\ifmmode
  \breve{g}\else \u{g}\fi{}mez-du Pree, Battiston, Behlmann, Beranek, Berdugo,
  Bertucci, Bindi, Bollweg, Borgia, Boschini, Bourquin, Bueno, Burger, Burger,
  Burmeister, Cai, Capell, Casaus, Castellini, Cervelli, Chang, Chen, Chen,
  Chen, Chen, Cheng, Chou, Chouridou, Choutko, Chung, Clark, Coignet,
  Consolandi, Contin, Corti, Cui, Dadzie, Dass, Delgado, Della~Torre,
  Demirk\"oz, Derome, Di~Falco, Di~Felice, D\'{\i}az, Dimiccoli, von
  Doetinchem, Dong, Donnini, Duranti, Egorov, Eline, Feng, Fiandrini, Fisher,
  Formato, Freeman, G\'amez, Garc\'{\i}a-L\'opez, Gargiulo, Gast, Gervasi,
  Giovacchini, G\'omez-Coral, Gong, Goy, Grabski, Grandi, Graziani, Haino, Han,
  Hashmani, He, Heber, Hsieh, Hu, Incagli, Jang, Jia, Jinchi, Karag\"oz,
  Khiali, Kim, Kirn, Konyushikhin, Kounina, Kounine, Koutsenko, Krasnopevtsev,
  Kuhlman, Kulemzin, La~Vacca, Laudi, Laurenti, Lazzizzera, Lebedev, Lee, Lee,
  Li, Li, Li, Li, Li, Li, Liang, Light, Lin, Lippert, Liu, Liu, Lu, Lu,
  Luebelsmeyer, Luo, Luo, Machate, Ma\~n\'a, Mar\'{\i}n, Marquardt, Martin,
  Mart\'{\i}nez, Masi, Maurin, Medvedeva, Menchaca-Rocha, Meng, Mikhailov,
  Molero, Mott, Mussolin, Negrete, Nikonov, Nozzoli, Oliva, Orcinha, Palermo,
  Palmonari, Paniccia, Pashnin, Pauluzzi, Pensotti, Phan, Plyaskin, Pohl,
  Poluianov, Qin, Qu, Quadrani, Rancoita, Rapin, Conde, Robyn, Rosier-Lees,
  Rozhkov, Rozza, Sagdeev, Schael, von Dratzig, Schwering, Seo, Shakfa, Shan,
  Siedenburg, Solano, Song, Song, Sonnabend, Strigari, Su, Sun, Sun, Tacconi,
  Tang, Tang, Tian, Ting, Ting, Tomassetti, Torsti, Urban, Usoskin, Vagelli,
  Vainio, Valencia-Otero, Valente, Valtonen, V\'azquez~Acosta, Vecchi, Velasco,
  Vialle, Wang, Wang, Wang, Wang, Wang, Wang, Wang, Wang, Wang, Wei, Weng, Wu,
  Xiong, Xu, Yan, Yang, Yashin, Yi, Yu, Yu, Zannoni, Zhang, Zhang, Zhang,
  Zhang, Zhang, Zhao, Zheng, Zheng, Zhuang, Zhukov, Zichichi, \&
  Zuccon}]{ams22p}
Aguilar, M., Cavasonza, L.~A., Ambrosi, G., {et~al.} 2021, Phys. Rev. Lett.,
  127, 271102

\bibitem[{Aguilar {et~al.}(2022)Aguilar, Cavasonza, Ambrosi, Arruda, Attig,
  Barao, Barrin, Bartoloni, Ba\ifmmode \mbox{\c{s}}\else \c{s}\fi{}e\ifmmode
  \breve{g}\else \u{g}\fi{}mez-du Pree, Battiston, Behlmann, Berdugo, Bertucci,
  Bindi, Bollweg, Borgia, Boschini, Bourquin, Bueno, Burger, Burger,
  Burmeister, Cai, Capell, Casaus, Castellini, Cervelli, Chang, Chen, Chen,
  Chen, Chen, Cheng, Chou, Chouridou, Choutko, Chung, Clark, Coignet,
  Consolandi, Contin, Corti, Cui, Dadzie, Dass, Delgado, Della~Torre,
  Demirk\"oz, Derome, Di~Falco, Di~Felice, D\'{\i}az, Dimiccoli, von
  Doetinchem, Dong, Donnini, Duranti, Egorov, Eline, Feng, Fiandrini, Fisher,
  Formato, Freeman, G\'amez, Garc\'{\i}a-L\'opez, Gargiulo, Gast, Gervasi,
  Giovacchini, G\'omez-Coral, Gong, Goy, Grabski, Grandi, Graziani, Haino, Han,
  Hashmani, He, Heber, Hsieh, Hu, Incagli, Jang, Jia, Jinchi, Karag\"oz,
  Khiali, Kim, Kirn, Konyushikhin, Kounina, Kounine, Koutsenko, Krasnopevtsev,
  Kuhlman, Kulemzin, La~Vacca, Laudi, Laurenti, Lazzizzera, Lee, Lee, Li, Li,
  Li, Li, Li, Li, Li, Li, Li, Liang, Liang, Light, Lin, Lippert, Liu, Lu, Lu,
  Luebelsmeyer, Luo, Luo, Machate, Ma\~n\'a, Mar\'{\i}n, Marquardt, Martin,
  Mart\'{\i}nez, Masi, Maurin, Medvedeva, Menchaca-Rocha, Meng, Mikhailov,
  Molero, Mott, Mussolin, Negrete, Nikonov, Nozzoli, Ocampo-Peleteiro, Oliva,
  Orcinha, Palermo, Palmonari, Paniccia, Pashnin, Pauluzzi, Pensotti, Plyaskin,
  Pohl, Poluianov, Qin, Qu, Quadrani, Rancoita, Rapin, Conde, Robyn,
  Rosier-Lees, Rozhkov, Rozza, Sagdeev, Schael, von Dratzig, Schwering, Seo,
  Shan, Siedenburg, Song, Song, Sonnabend, Strigari, Su, Sun, Sun, Tacconi,
  Tang, Tang, Tian, Ting, Ting, Tomassetti, Torsti, Urban, Usoskin, Vagelli,
  Vainio, Valencia-Otero, Valente, Valtonen, V\'azquez~Acosta, Vecchi, Velasco,
  Vialle, Wang, Wang, Wang, Wang, Wang, Wang, Wang, Wang, Wang, Wei, Weng, Wu,
  Xiong, Xu, Yan, Yang, Yashin, Yi, Yu, Yu, Zannoni, Zhang, Zhang, Zhang,
  Zhang, Zhang, Zhao, Zheng, Zheng, Zhuang, Zhukov, Zichichi, \&
  Zuccon}]{ams22he}
Aguilar, M., Cavasonza, L.~A., Ambrosi, G., {et~al.} 2022, Phys. Rev. Lett.,
  128, 231102

\bibitem[{{Andretta} {et~al.}(2014){Andretta}, {Bemporad}, {Focardi},
  {Grimani}, {Landini}, {Pancrazzi}, {Sasso}, {Spadaro}, {Straus}, {Uslenghi},
  {Antonucci}, {Fineschi}, {Naletto}, {Nicolini}, {Nicolosi}, \&
  {Romoli}}]{andre}
{Andretta}, V., {Bemporad}, A., {Focardi}, M., {et~al.} 2014, in Society of
  Photo-Optical Instrumentation Engineers (SPIE) Conference Series, Vol. 9152,
  Software and Cyberinfrastructure for Astronomy III, 91522Q

\bibitem[{Antonucci {et~al.}(2023)Antonucci, Downs, Capuano, Spadaro, Susino,
  Telloni, Andretta, Da~Deppo, De~Leo, Fineschi, Frassetto, Landini, Naletto,
  Nicolini, Pancrazzi, Romoli, Stangalini, Teriaca, \& Uslenghi}]{antonucci23}
Antonucci, E., Downs, C., Capuano, G.~E., {et~al.} 2023, Physics of Plasmas,
  30, 022905

\bibitem[{{Antonucci} {et~al.}(2020){Antonucci}, {Romoli}, {Andretta},
  {Fineschi}, {Heinzel, Petr}, {Moses, J. Daniel}, {Naletto, Giampiero},
  {Nicolini, Gianalfredo}, {Spadaro, Daniele}, {Teriaca, Luca}, {Berlicki,
  Arkadiusz}, {Capobianco, Gerardo}, {Crescenzio, Giuseppe}, {Da Deppo, Vania},
  {Focardi, Mauro}, {Frassetto, Fabio}, {Heerlein, Klaus}, {Landini, Federico},
  {Magli, Enrico}, {Marco Malvezzi, Andrea}, {Massone, Giuseppe}, {Melich,
  Radek}, {Nicolosi, Piergiorgio}, {Noci, Giancarlo}, {Pancrazzi, Maurizio},
  {Pelizzo, Maria G.}, {Poletto, Luca}, {Sasso, Clementina}, {Sch\"uhle, Udo},
  {Solanki, Sami K.}, {Strachan, Leonard}, {Susino, Roberto}, {Tondello,
  Giuseppe}, {Uslenghi, Michela}, {Woch, Joachim}, {Abbo, Lucia}, {Bemporad,
  Alessandro}, {Casti, Marta}, {Dolei, Sergio}, {Grimani, Catia}, {Messerotti,
  Mauro}, {Ricci, Marco}, {Straus, Thomas}, {Telloni, Daniele}, {Zuppella,
  Paola}, {Auch\`ere, Frederic}, {Bruno, Roberto}, {Ciaravella, Angela},
  {Corso, Alain J.}, {Alvarez Copano, Miguel}, {Aznar Cuadrado, Regina},
  {D\'{}Amicis, Raffaella}, {Enge, Reiner}, {Gravina, Alessio}, {Jejcic,
  Sonja}, {Lamy, Philippe}, {Lanzafame, Alessandro}, {Meierdierks, Thimo},
  {Papagiannaki, Ioanna}, {Peter, Hardi}, {Fernandez Rico, German}, {Giday
  Sertsu, Mewael}, {Staub, Jan}, {Tsinganos, Kanaris}, {Velli, Marco},
  {Ventura, Rita}, {Verroi, Enrico}, {Vial, Jean-Claude}, {Vives, Sebastien},
  {Volpicelli, Antonio}, {Werner, Stephan}, {Zerr, Andreas}, {Negri, Barbara},
  {Castronuovo, Marco}, {Gabrielli, Alessandro}, {Bertacin, Roberto},
  {Carpentiero, Rita}, {Natalucci, Silvia}, {Marliani, Filippo}, {Cesa, Marco},
  {Laget, Philippe}, {Morea, Danilo}, {Pieraccini, Stefano}, {Radaelli, Paolo},
  {Sandri, Paolo}, {Sarra, Paolo}, {Cesare, Stefano}, {Del Forno, Felice},
  {Massa, Ernesto}, {Montabone, Mauro}, {Mottini, Sergio}, {Quattropani,
  Daniele}, {Schillaci, Tiziano}, {Boccardo, Roberto}, {Brando, Rosario},
  {Pandi, Arianna}, {Baietto, Cristian}, {Bertone, Riccardo}, {Alvarez-Herrero,
  Alberto}, {Garc\'{\i}a Parejo, Pilar}, {Cebollero, Mar\'{\i}a}, {Amoruso,
  Mauro}, \& {Centonze, Vito}}]{a&ametis}
{Antonucci}, E., {Romoli}, M., {Andretta}, V., {et~al.} 2020, A\&A, 642, A10

\bibitem[{{Armano} {et~al.}(2018){Armano}, {Audley}, {Baird}, {Bassan},
  {Benella}, {Binetruy}, {Born}, {Bortoluzzi}, {Cavalleri}, {Cesarini},
  {Cruise}, {Danzmann}, {de Deus Silva}, {Diepholz}, {Dixon}, {Dolesi}, {Fabi},
  {Ferraioli}, {Ferroni}, {Finetti}, {Fitzsimons}, {Freschi}, {Gesa}, {Gibert},
  {Giardini}, {Giusteri}, {Grimani}, {Grzymisch}, {Harrison}, {Heinzel},
  {Hewitson}, {Hollington}, {Hoyland}, {Hueller}, {Inchausp{\'e}}, {Jennrich},
  {Jetzer}, {Karnesis}, {Kaune}, {Korsakova}, {Killow}, {Laurenza}, {Lobo},
  {Lloro}, {Liu}, {L{\'o}pez-Zaragoza}, {Maarschalkerweerd}, {Mance},
  {Mart{\'{\i}}n}, {Martin-Polo}, {Martino}, {Martin-Porqueras}, {Mateos},
  {McNamara}, {Mendes}, {Mendes}, {Nofrarias}, {Paczkowski}, {Perreur-Lloyd},
  {Petiteau}, {Pivato}, {Plagnol}, {Ramos-Castro}, {Reiche}, {Robertson},
  {Rivas}, {Russano}, {Sabbatini}, {Slutsky}, {Sopuerta}, {Sumner}, {Telloni},
  {Texier}, {Thorpe}, {Vetrugno}, {Vitale}, {Wanner}, {Ward}, {Wass}, {Weber},
  {Wissel}, {Wittchen}, {Zambotti}, {Zenoni}, \& {Zweifel}}]{apj1}
{Armano}, M., {Audley}, H., {Baird}, J., {et~al.} 2018, Astrophys. J., 854, 113

\bibitem[{{Battistoni} {et~al.}(2014){Battistoni}, {Boehlen}, {Cerutti},
  {Chin}, {Salvatore Esposito}, {Fass{\`o}}, {Ferrari}, {Mereghetti}, {Garcia
  Ortega}, {Ranft}, {Roesler}, {Sala}, \& {Vlachoudis}}]{flukacern1}
{Battistoni}, G., {Boehlen}, T., {Cerutti}, F., {et~al.} 2014, in Joint
  International Conference on Supercomputing in Nuclear Applications + Monte
  Carlo, 06005

\bibitem[{{B{\"o}hlen} {et~al.}(2014){B{\"o}hlen}, {Cerutti}, {Chin},
  {Fass{\`o}}, {Ferrari}, {Ortega}, {Mairani}, {Sala}, {Smirnov}, \&
  {Vlachoudis}}]{flukacern2}
{B{\"o}hlen}, T.~T., {Cerutti}, F., {Chin}, M.~P.~W., {et~al.} 2014, Nuclear
  Data Sheets, 120, 211

\bibitem[{{Brehm} {et~al.}(2021){Brehm}, {Bayliss}, {Christl}, {Synal},
  {Adolphi}, {Beer}, {Kromer}, {Muscheler}, {Solanki}, {Usoskin}, {Bleicher},
  {Bollhalder}, {Tyers}, \& {Wacker}}]{brehm21}
{Brehm}, N., {Bayliss}, A., {Christl}, M., {et~al.} 2021, Nature Geoscience,
  14, 10

\bibitem[{Burger {et~al.}(2000)Burger, Potgieter, \& Heber}]{burger2000}
Burger, R.~A., Potgieter, M.~S., \& Heber, B. 2000, Journal of Geophysical
  Research: Space Physics, 105, 27447

\bibitem[{{Clette} {et~al.}(2014){Clette}, {Svalgaard}, {Vaquero}, \&
  {Cliver}}]{silso}
{Clette}, F., {Svalgaard}, L., {Vaquero}, J.~M., \& {Cliver}, E.~W. 2014, \ssr,
  186, 35

\bibitem[{Dalla {et~al.}(2003)Dalla, Balogh, Krucker, Posner, Müller-Mellin,
  Anglin, Hofer, Marsden, Sanderson, Tranquille, Heber, Zhang, \&
  McKibben}]{dalla03}
Dalla, S., Balogh, A., Krucker, S., {et~al.} 2003, Geophysical Research
  Letters, 30
  [\eprint{https://agupubs.onlinelibrary.wiley.com/doi/pdf/10.1029/2003GL017139}]

\bibitem[{{De Simone} {et~al.}(2011){De Simone}, {Di Felice}, {Gieseler},
  {Boezio}, {Casolino}, {Picozza}, {Heber}, \& {PAMELA
  Collaboration}}]{desimone}
{De Simone}, N., {Di Felice}, V., {Gieseler}, J., {et~al.} 2011, Astrophysics
  and Space Sciences Transactions, 7, 425

\bibitem[{{Fineschi} {et~al.}(2020){Fineschi}, {Naletto}, {Romoli}, {Da Deppo},
  {Antonucci}, {Moses}, {Malvezzi}, {Nicolini}, {Spadaro}, {Teriaca},
  {Andretta}, {Capobianco}, {Crescenzio}, {Focardi}, {Frassetto}, {Landini},
  {Massone}, {Melich}, {Nicolosi}, {Pancrazzi}, {Pelizzo}, {Poletto},
  {Sch{\"u}hle}, {Uslenghi}, {Vives}, {Solanki}, {Heinzel}, {Berlicki},
  {Cesare}, {Morea}, {Mottini}, {Sandri}, {Alvarez-Herrero}, \&
  {Castronuovo}}]{fineschi2020}
{Fineschi}, S., {Naletto}, G., {Romoli}, M., {et~al.} 2020, Experimental
  Astronomy, 49, 239

\bibitem[{{Garc\'{\i}a Marirrodriga} {et~al.}(2021){Garc\'{\i}a Marirrodriga},
  {Pacros}, {Strandmoe, S.}, {Arcioni, M.}, {Arts, A.}, {Ashcroft, C.},
  {Ayache, L.}, {Bonnefous, Y.}, {Brahimi, N.}, {Cipriani, F.}, {Damasio, C.},
  {De Jong, P.}, {D\'eprez, G.}, {Fahmy, S.}, {Fels, R.}, {Fiebrich, J.},
  {Hass, C.}, {Hern\'andez, C.}, {Icardi, L.}, {Junge, A.}, {Kletzkine, P.},
  {Laget, P.}, {Le Deuff, Y.}, {Liebold, F.}, {Lodiot, S.}, {Marliani, F.},
  {Mascarello, M.}, {M\"uller, D.}, {Oganessian, A.}, {Olivier, P.}, {Palombo,
  E.}, {Philippe, C.}, {Ragnit, U.}, {Ramachandran, J.}, {S\'anchez P\'erez, J.
  M.}, {Stienstra, M. M.}, {Th\"urey, S.}, {Urwin, A.}, {Wirth, K.}, \&
  {Zouganelis, I.}}]{a&asolomission}
{Garc\'{\i}a Marirrodriga}, C., {Pacros}, A., {Strandmoe, S.}, {et~al.} 2021,
  A\&A, 646, A121

\bibitem[{Gleeson \& Axford(1968)}]{glax68}
Gleeson, L.~J. \& Axford, W.~I. 1968, Ap. J., 154, 1011

\bibitem[{{Grimani}(2004)}]{gri04}
{Grimani}, C. 2004, \aap, 418, 649

\bibitem[{{Grimani}(2007)}]{gri07}
{Grimani}, C. 2007, \aap, 474, 339

\bibitem[{{Grimani} {et~al.}(2021){Grimani}, {Andretta}, {Chioetto}, {Da
  Deppo}, {Fabi}, {Gissot}, {Naletto}, {Persici}, {Plainaki}, {Romoli},
  {Sabbatini}, {Spadaro}, {Stangalini}, {Telloni}, {Uslenghi}, {Antonucci},
  {Bemporad}, {Capobianco}, {Capuano}, {Casti}, {De Leo}, {Fineschi},
  {Frassati}, {Frassetto}, {Heinzel}, {Jerse}, {Landini}, {Liberatore},
  {Magli}, {Messerotti}, {Moses}, {Nicolini}, {Pancrazzi}, {Pelizzo}, {Romano},
  {Sasso}, {Sch{\"u}hle}, {Slemer}, {Straus}, {Susino}, {Teriaca},
  {Volpicelli}, {Freiherr von Forstner}, \& {Zuppella}}]{a&aub}
{Grimani}, C., {Andretta}, V., {Chioetto}, P., {et~al.} 2021, \aap, 656, A15

\bibitem[{{Grimani} {et~al.}(2013){Grimani}, {Fabi}, {Finetti}, {Laurenza}, \&
  {Storini}}]{spectral}
{Grimani}, C., {Fabi}, M., {Finetti}, N., {Laurenza}, M., \& {Storini}, M.
  2013, in Journal of Physics Conference Series, Vol. 409, Journal of Physics
  Conference Series, 012159

\bibitem[{{Grimani} {et~al.}(2008){Grimani}, {Fabi}, {Finetti}, \&
  {Tombolato}}]{grim07}
{Grimani}, C., {Fabi}, M., {Finetti}, N., \& {Tombolato}, D. 2008,
  International Cosmic Ray Conference, 1, 485

\bibitem[{Grimani {et~al.}(2009)Grimani, Fabi, Finetti, \& Tombolato}]{griele}
Grimani, C., Fabi, M., Finetti, N., \& Tombolato, D. 2009, Class. Quant. Grav.,
  26, 215004

\bibitem[{{Grimani} {et~al.}(2022){Grimani}, {Villani}, {Fabi}, {Cesarini}, \&
  {Sabbatini}}]{bridge}
{Grimani}, C., {Villani}, M., {Fabi}, M., {Cesarini}, A., \& {Sabbatini}, F.
  2022, \aap, 666, A38

\bibitem[{Lechner {et~al.}(2019)Lechner, Auchmann, Baer, Bahamonde~Castro,
  Bruce, Cerutti, Esposito, Ferrari, Jowett, Mereghetti, Pietropaolo, Redaelli,
  Salvachua, Sapinski, Schaumann, Shetty, Vlachoudis, \& Skordis}]{Lechner19}
Lechner, A., Auchmann, B., Baer, T., {et~al.} 2019, Phys. Rev. Accel. Beams,
  22, 071003

\bibitem[{{Marquardt} \& {Heber}(2019)}]{helios6}
{Marquardt}, J. \& {Heber}, B. 2019, A\&A, 625, A153

\bibitem[{Martucci {et~al.}(2023)Martucci, Laurenza, Benella, Berrilli,
  Del~Moro, Giovannelli, Parmentier, Piersanti, Albrecht, Bartocci, Battiston,
  Burger, Campana, Carfora, Consolini, Conti, Contin, De~Donato, De~Santis,
  Follega, Iuppa, Lega, Marcelli, Masciantonio, Mergé, Mese, Oliva, Osteria,
  Palma, Panico, Perfetto, Picozza, Pozzato, Ricci, Ricci, Ricciarini, Sahnoun,
  Scotti, Sotgiu, Sparvoli, Vitale, Zoffoli, \& Zuccon}]{martucci23}
Martucci, M., Laurenza, M., Benella, S., {et~al.} 2023, Space Weather, 21,
  e2022SW003191, e2022SW003191 2022SW003191

\bibitem[{{M\"uller} {et~al.}(2020){M\"uller}, {St. Cyr}, {Zouganelis},
  {Gilbert}, {Marsden}, {Nieves-Chinchilla, T.}, {Antonucci, E.}, {Auch\`ere,
  F.}, {Berghmans, D.}, {Horbury, T. S.}, {Howard, R. A.}, {Krucker, S.},
  {Maksimovic, M.}, {Owen, C. J.}, {Rochus, P.}, {Rodriguez-Pacheco, J.},
  {Romoli, M.}, {Solanki, S. K.}, {Bruno, R.}, {Carlsson, M.}, {Fludra, A.},
  {Harra, L.}, {Hassler, D. M.}, {Livi, S.}, {Louarn, P.}, {Peter, H.},
  {Sch\"uhle, U.}, {Teriaca, L.}, {del Toro Iniesta, J. C.},
  {Wimmer-Schweingruber, R. F.}, {Marsch, E.}, {Velli, M.}, {De Groof, A.},
  {Walsh, A.}, \& {Williams, D.}}]{a&asoloscience}
{M\"uller}, D., {St. Cyr}, O.~C., {Zouganelis}, I., {et~al.} 2020, A\&A, 642,
  A1

\bibitem[{Nymmik(1999a)}]{nym99a}
Nymmik, R. 1999a, in 26th Int. Cosmic Ray Conf. (Salt Lake City), Vol.~6,
  268--271

\bibitem[{Nymmik(1999b)}]{nym99b}
Nymmik, R. 1999b, in 26th Int. Cosmic Ray Conf. (Salt Lake City), Vol.~6,
  280--283

\bibitem[{{Papaioannou} {et~al.}(2022){Papaioannou}, {Kouloumvakos}, {Mishev},
  {Vainio}, {Usoskin}, {Herbst}, {Rouillard}, {Anastasiadis}, {Gieseler},
  {Wimmer-Schweingruber}, \& {K{\"u}hl}}]{papa22}
{Papaioannou}, A., {Kouloumvakos}, A., {Mishev}, A., {et~al.} 2022, \aap, 660,
  L5

\bibitem[{Papini {et~al.}(1996)Papini, Grimani, \& Stephens}]{papini96}
Papini, P., Grimani, C., \& Stephens, S. 1996, Nuovo Cim., C19, 367

\bibitem[{{Reames}(2021)}]{reames21}
{Reames}, D.~V. 2021, {Solar Energetic Particles. A Modern Primer on
  Understanding Sources, Acceleration and Propagation}, Vol. 978

\bibitem[{{Romoli} {et~al.}(2021){Romoli}, {Antonucci}, {Andretta}, {Capuano},
  {Da Deppo}, {De Leo}, {Downs}, {Fineschi}, {Heinzel}, {Landini},
  {Liberatore}, {Naletto}, {Nicolini}, {Pancrazzi}, {Sasso}, {Spadaro},
  {Susino}, {Telloni}, {Teriaca}, {Uslenghi}, {Wang}, {Bemporad}, {Capobianco},
  {Casti}, {Fabi}, {Frassati}, {Frassetto}, {Giordano}, {Grimani}, {Jerse},
  {Magli}, {Massone}, {Messerotti}, {Moses}, {Pelizzo}, {Romano},
  {Sch{\"u}hle}, {Slemer}, {Stangalini}, {Straus}, {Volpicelli}, {Zangrilli},
  {Zuppella}, {Abbo}, {Auch{\`e}re}, {Aznar Cuadrado}, {Berlicki}, {Bruno},
  {Ciaravella}, {D'Amicis}, {Lamy}, {Lanzafame}, {Malvezzi}, {Nicolosi},
  {Nistic{\`o}}, {Peter}, {Plainaki}, {Poletto}, {Reale}, {Solanki},
  {Strachan}, {Tondello}, {Tsinganos}, {Velli}, {Ventura}, {Vial}, {Woch}, \&
  {Zimbardo}}]{a&ametis2}
{Romoli}, M., {Antonucci}, E., {Andretta}, V., {et~al.} 2021, \aap, 656, A32

\bibitem[{{Sch{\"u}hle} {et~al.}(2018){Sch{\"u}hle}, {Teriaca}, {Aznar
  Cuadrado}, {Heerlein}, {Uslenghi}, \& {Werner}}]{udo18}
{Sch{\"u}hle}, U., {Teriaca}, L., {Aznar Cuadrado}, R., {et~al.} 2018, in
  Society of Photo-Optical Instrumentation Engineers (SPIE) Conference Series,
  Vol. 10699, Space Telescopes and Instrumentation 2018: Ultraviolet to Gamma
  Ray, ed. J.-W.~A. {den Herder}, S.~{Nikzad}, \& K.~{Nakazawa}, 1069934

\bibitem[{Shikaze {et~al.}(2007)Shikaze, Haino, Abe, Fuke, Hams, Kim, Makida,
  Matsuda, Mitchell, Moiseev, Nishimura, Nozaki, Orito, Ormes, Sanuki, Sasaki,
  Seo, Streitmatter, Suzuki, Tanaka, Yamagami, Yamamoto, Yoshida, \&
  Yoshimura}]{Shikaze2007154}
Shikaze, Y., Haino, S., Abe, K., {et~al.} 2007, Astroparticle Physics, 28, 154

\bibitem[{{Singh} \& {Bhargawa}(2019)}]{singh19}
{Singh}, A.~K. \& {Bhargawa}, A. 2019, \apss, 364, 12

\bibitem[{Sullivan(1971)}]{sullivan}
Sullivan, J. 1971, Nuclear Instruments and Methods, 95, 5

\bibitem[{Telloni {et~al.}(2016)Telloni, Fabi, Grimani, \&
  Antonucci}]{solwind14}
Telloni, D., Fabi, M., Grimani, C., \& Antonucci, E. 2016, AIP Conference
  Proceedings, 1720

\bibitem[{{Telloni} {et~al.}(2022){Telloni}, {Zank}, {Stangalini}, {Downs},
  {Liang}, {Nakanotani}, {Andretta}, {Antonucci}, {Sorriso-Valvo}, {Adhikari},
  {Zhao}, {Marino}, {Susino}, {Grimani}, {Fabi}, {D'Amicis}, {Perrone},
  {Bruno}, {Carbone}, {Mancuso}, {Romoli}, {Deppo}, {Fineschi}, {Heinzel},
  {Moses}, {Naletto}, {Nicolini}, {Spadaro}, {Teriaca}, {Frassati}, {Jerse},
  {Landini}, {Pancrazzi}, {Russano}, {Sasso}, {Biondo}, {Burtovoi}, {Capuano},
  {Casini}, {Casti}, {Chioetto}, {Leo}, {Giarrusso}, {Liberatore}, {Berghmans},
  {Auch{\`e}re}, {Cuadrado}, {Chitta}, {Harra}, {Kraaikamp}, {Long}, {Mandal},
  {Parenti}, {Pelouze}, {Peter}, {Rodriguez}, {Sch{\"u}hle}, {Schwanitz},
  {Smith}, {Verbeeck}, \& {Zhukov}}]{tellonisw}
{Telloni}, D., {Zank}, G.~P., {Stangalini}, M., {et~al.} 2022, \apjl, 936, L25

\bibitem[{{Uslenghi} {et~al.}(2017){Uslenghi}, {Sch{\"u}hle}, {Teriaca},
  {Heerlein}, \& {Werner}}]{uslenghi2017}
{Uslenghi}, M., {Sch{\"u}hle}, U.~H., {Teriaca}, L., {Heerlein}, K., \&
  {Werner}, S. 2017, in Society of Photo-Optical Instrumentation Engineers
  (SPIE) Conference Series, Vol. 10397, Society of Photo-Optical
  Instrumentation Engineers (SPIE) Conference Series, ed. O.~H. {Siegmund},
  103971K

\bibitem[{{Usoskin} {et~al.}(2011){Usoskin}, {Bazilevskaya}, \&
  {Kovaltsov}}]{uso11}
{Usoskin}, I.~G., {Bazilevskaya}, G.~A., \& {Kovaltsov}, G.~A. 2011, Journal of
  Geophysical Research (Space Physics), 116, A02104

\bibitem[{{Usoskin} {et~al.}(2017){Usoskin}, {Gil}, {Kovaltsov}, {Mishev}, \&
  {Mikhailov}}]{uso17}
{Usoskin}, I.~G., {Gil}, A., {Kovaltsov}, G.~A., {Mishev}, A.~L., \&
  {Mikhailov}, V.~V. 2017, Journal of Geophysical Research (Space Physics),
  122, 3875

\bibitem[{Vlachoudis(2009)}]{flair}
Vlachoudis, V. 2009, in {International Conference on Mathematics, Computational
  Methods \& Reactor Physics (M\&C 2009), Saratoga Springs, New York}, 790--800

\bibitem[{{Wimmer-Schweingruber} {et~al.}(2021){Wimmer-Schweingruber},
  {Pacheco}, {Janitzek}, {Cernuda}, {Espinosa Lara}, {Gómez-Herrero}, {Mason},
  {Allen}, {Xu}, {Carcaboso}, {Kollhoff}, {Kühl}, {Freiherr von Forstner},
  {Berger}, {Rodriguez-Pacheco}, {Ho}, {Andrews}, {Angelini}, {Aran}, {Boden},
  {Böttcher}, {Carrasco}, {Dresing}, {Eldrum}, {Elftmann}, {Evans}, {Gevin},
  {Hayes}, {Heber}, {Horbury}, {Kulkarni}, {Lario}, {Lees}, {Limousin},
  {Malandraki}, {Martín}, {O’Brien}, {Prieto Mateo}, {Ravanbakhsh},
  {Rodriguez Polo}, {Sánchez Prieto}, {Schlemm}, {Seifert}, {Terasa },
  {Tyagi}, {Vainio}, {Walsh}, \& {Yedla}}]{wimmer21}
{Wimmer-Schweingruber}, R.~F., {Pacheco}, D., {Janitzek}, N., {et~al.} 2021,
  submitted to A\&A

\end{thebibliography}
%

\end{document}